**Quantitative Phase Imaging with a Metalens**

*Aamod Shanker,[1,4], Johannes Fröch[1], Saswata Mukherjee, Maksym Zhelyeznyakov[1], Eric Seibel[3], Arka Majumdar[1,2]*

[1]Department of Electrical and Computer Engineering, University of Washington, Seattle, WA, 98195, USA

[2]Department of Physics, University of Washington, Seattle, WA-98195 USA

[3]Department of Mechanical Engineering, University of Washington, Seattle, WA 98105, USA

[4]Center for Vision Science, University of Rochester, New York, NY 14623, USA

* Corresponding author: arka@uw.edu

**Abstract**

Quantitative phase imaging (QPI) recovers the exact wavefront of light from the intensity measured by a camera. Topographical maps of translucent microscopic bodies can be extracted from these quantified phase shifts. We demonstrate quantitative phase imaging at the tip of an optical fiber endoscope with a chromatic silicon nitride metalens. Our method leverages spectral multiplexing to recover phase from multiple defocus planes in a single capture. The half millimeter wide metalens shows phase imaging capability with a $28^0$ field of view and $0.1\lambda$ sensitivity in experiments with an endoscopic fiber bundle. Since the spectral functionality is encoded directly in the imaging lens, no additional filters are needed. Key limitations in the scaling of a phase imaging system, such as multiple acquisition, interferometric alignment or mechanical scanning are completely mitigated in the proposed scheme.

**Introduction**

Thanks to the progress in nanofabrication technology, availability of advanced electromagnetic simulators, and presence of computing elements, imaging systems have been drastically miniaturized. On one hand, it is now possible to create sub-wavelength flat diffractive optics, commonly known as meta-optics, with large phase-gradients[1-4]. On the other hand, sophisticated computational imaging techniques can process the captured data to extract new information without adding a significant footprint[5,6]. Such miniaturization of imaging systems and sensors goes hand-in-hand with emerging applications, such as autonomous systems and internet of things[7]. Biomedical imaging systems such as endoscopes[8,9] or angioscopes are yet another application space which necessitates continuous miniaturization of optical elements. Imaging systems at ever diminishing scales can benefit from capturing multi-dimensional data, and often require information beyond pure intensity measurement. Specifically, phase information is often needed for semi-transparent biomedical tissues.

Quantitative Phase Imaging (QPI) microscopy [10] is able to extract precise quantitative information about optical path length from optical intensity [11]. Knowing both the intensity and phase allows for 3D imaging and reconstruction by digitally back-propagating the light path [12], or 3D beam steering for computer generated holography [13]. Applications of QPI have vastly expanded since it was first proposed as a non-interferometric partially coherent phase imaging method [14,15], including X-Ray phase imaging [16,17], atomic and molecular structure imaging, or polarization dependent mask edge effects in lithography [18]. The ability to extract morphological information from quantitative phase without biomarkers or fluorescent labels enables characterization of cell structure in its native living state. Especially attractive in-vivo applications are non-invasive diagnostic techniques in ophthalmoscopy [19,20], as well as epithelial cell imaging in the gastrointestinal tract or intra-organ cavities with endoscopy[21]. However, quantitative phase imaging of light-tissue interaction in a single-shot capture for real time endoscopy is yet to be achieved.

Conventional phase imaging methods are difficult to translate to in-vivo applications at ophthalmological (1-20mm) or endoscopic (<2mm) scales due to fine alignment requirements in interference-based measurements between a reference beam and scattered beam. Additional mechanical complexity is introduced in scanning methods such as optical coherence tomography or confocal microscopy [22], where point scanning needs moving mirrors that have to simultaneously aligned spatially and temporally. Interferometric phase measurement systems also often have strict coherence requirements necessitating the use of tunable laser systems. Hence interferometric or point scanning methods need complex actuators which hinder translation and adoption in clinical applications due to low volume and high product cost. While lens-less adoptions of endoscopy can mitigate these problems, they suffer from speckle and honeycomb artifacts or calibration sensitivity where fiber bending ruins the reconstruction [23].

In this paper, we demonstrate a single-shot QPI method that utilizes the inherent chromaticity of metaoptics. Metalenses, like any other diffractive optics, suffer from strong axial chromatic aberration, with a wavelength-dependent focal length following a relation: $\delta f/f = -\delta\lambda/\lambda$, given $\lambda f = constant$[24]. The chromatic aberration causes colors ($\delta\lambda$) to image at different focal distances ($\delta f$), creating artifacts such as color vignetting at sharp features and edges. In recent years, researchers have developed many techniques to mitigate the chromatic aberrations in a meta-lens [3,5,6,25]. Here, we harness the chromatic aberration for quantitative phase imaging ("QPI"), especially in the context of endoscopic applications where phase is scrambled on passage through fiber cores. Some recent works have demonstrated three dimensional imaging using meta-optics, although quantitative phase retrieval through a fiber bundle is yet to be demonstrated [26,27]. Others have shown quantitative phase imaging using complex meta-optical designs unsuitable for fiber endoscopy[28] or needing elaborate end to end phase calibration for each fiber in a fiber bundle [29]. Our single-shot QPI approach overcomes alignment or acquisition limitations of typical phase imaging methods by extracting the whole electric field from a single bright-field color image. We demonstrate robust phase retrieval with commercial-off-the-shelf light-emitting diodes and cameras at real-time acquisition speeds. We further demonstrate the ability to decipher structural features from spectral information in meta-optical endoscopy. We finally demonstrate QPI through a coherent fiber bundle. The principles utilized here have a high potential for impact in the medical diagnostic setting with sub-mm apertures, especially when transmitting through an optical fiber or other media that preserves spectral intensity $I(\lambda_r, \lambda_g, \lambda_b)$ while scrambling coherent phase $\phi$.

**Results**

Inspired by the focal length dispersion in a meta-lens, we have developed a volumetric imaging technique that encodes depth into spectrum. The method is non-interferometric, since the phase is computed from the multiple defocus measurements; the multiple focal planes are encoded in the RGB color channels in a single measurement [30]. The red, green and blue channels are centered around 455nm, 530nm and 625nm respectively, determined by the spectral response of the Bayer filter in the camera to the white light LED excitation (see Methods) [31]. Our design principles draw largely from the conjugate principle of light, which assumes the light path to be reversible, i.e., focusing is equivalent to imaging in reverse.

**QPI with the Transport of Intensity Equation**

A partially coherent light passing through transparent scatterers (disordered/ordered, organic /inorganic) encodes nanometer scale height variations into the phase of the wavefront as $\phi = \frac{2\pi}{\lambda} \times h \times \delta n$, where $h$ is the sample height and $\delta n$ is the refractive index difference between the sample and the surrounding media. The effective phase picked up during the propagation through the sample is encoded into the complex valued electric field $\bar{E} = \sqrt{I(x,y)} e^{i\phi(x,y)}$, where $I(x,y)$ is the 2D intensity that can be measured on a camera. For a linear, time-invariant system the electric field after propagation through a distance $z$ is given by the convolution of the input field with the paraxial point spread function (PSF): $h_z(x,y) = \frac{e^{ikz}}{i\lambda z} e^{i\frac{2\pi}{\lambda} \frac{x^2+y^2}{2z}}$. A further paraxial approximation that assumes propagation to be mostly along the z axis yields two complementary equations [32,33], (see Supplement S1 for full derivation).

*Transport of Intensity Equation (TIE):*
$$\frac{dI}{dz} = -\frac{\lambda}{2\pi} \widehat{\nabla} \cdot I \widehat{\nabla} \phi$$

*Transport of Phase Equation (TPE):*
$$\frac{2\pi}{\lambda} \frac{d\phi}{dz} = -\widehat{\nabla}\phi \cdot \widehat{\nabla}\phi + \frac{\nabla^2 I}{2I} - \frac{(\widehat{\nabla}I)^2}{4I^2}$$

Where, $\widehat{\nabla} = \frac{d}{dx}\hat{x} + \frac{d}{dy}\hat{y}$ is the *in-plane* gradient normal to the propagation direction. Together the two equations completely describe coherent propagation as an alternating update between phase and intensity at each defocus step. The Transport of Intensity Equation is similar to the continuity equation for fluids, with light Intensity substituting for fluid density. The Transport of Phase equation, on the other hand, is analogous to the momentum transport equation for fluids, where the time evolution of the fluid velocity is substituted by gradients of light phase. (See S1).

Additionally, for phase retrieval from spectral intensity measurement ($d\lambda$), we rewrite the Transport of Intensity [30] assuming fixed z and varying $\lambda$ as $\frac{dI}{d\lambda} = -\frac{z}{2\pi} \widehat{\nabla} \cdot I \widehat{\nabla} \phi$.

Reframing the intensity change as a function of chromatic wavelength ($dI/d\lambda$) at constant defocus distnace z describes the chromatic dispersion of the diffracted light at a certain propagation distance after scattering off an object with phase shift $\phi(x,y)$ and absorption $I(x,y)$. Hence, free space propagation is equivalent to spectral shift according to the transport equation for intensity, which effectively treats the variable $\zeta = \lambda z$ as a single combined variable. (confirmed experimentally in Fig 1).

**Metalens Design**

We now describe how a simple metalens design also preserves the conjugacy between focal distance and wavelength (Fig. 1). The metalens is specifically designed as a spectral diffraction grating along the longitudinal propagation direction *z*, effectively having linear longitudinal chromatic aberration (LCA). The metalens focal position has a spectral dependence where focal length and wavelength have an exact inverse-linear relationship. This chromatic dependence of focus is typically a nuisance; however, here we utilize the spectral-focal equivalence to encode multiple focal planes simultaneously into spectral channels. The lens is designed according to the hyperboloid phase profile $\phi(r) = \frac{2\pi}{\lambda}\frac{r^2}{2f}$ [32], to focus the green channel ($\lambda$ =530nm), with the red (625nm) and blue (445nm) channels automatically focusing ahead and behind the green respectively (Fig 1a, d) such that $\lambda z = const$ (Fig 1b). Consequently $\delta f/f = -\delta\lambda/\lambda$ ; hence he expected focal shift (between red and green ($\delta\lambda = 95nm,\ \lambda = 530nm,\ f = 1mm$) can be expected to be $\delta f = 180um$.

The measured $\lambda f$ values are nearly identical for the three colors, where f is the focal distance for the given $\lambda$ (see S5). We also measure $\delta f = 180um$ change in focal distance between the color channels, almost identical to the predicted value (Fig. 1d). Lenses with shorter focal lengths tend to have lesser chromatic aberration, since $\delta f = -f\ \delta\lambda/\lambda$. Hence shorter focal length lenses at longer wavelengths have reduced LCA . Additionally the negative sign implies that increasing the wavelength reduces the focal length, similar to other diffractive optics. Hence the longitudinal dispersion in our metalens causes red to focus before blue, similar to a diffractive lens, but opposite to that of a refractive lens [6].

We choose a nominal focal length of 1mm for the 0.5mm aperture hyperboloid metalens, with an effective Numerical Aperture (NA) of ~0.25. The objective used to relay the metalens image has a larger NA of 0.75, hence being able to capture the entire angular bandwidth relayed from the metalens. The metalens is made of SiN, with the rectangular post scatterers are designed by rigorous coupled wave analysis (See Supplement S5). Standard e-beam lithography followed by etching is used to fabricate the meta-optics. We characterized the focusing behavior of the metalens for three color channels (Fig. 1) and measured ~20% change in the focal length between red and blue channel. Hence, we can use the spectral channels as a surrogate for focal shift in our single shot version of quantitative phase imaging. Here, we utilize the three color channels in a camera's Bayer filter instead of taking three defocus measurements on a mechanical stage. The phase or wavefront can be computed by inverting the previously introduced transport equation for intensity that relates phase to defocus z and hence to color $\lambda$ [33]. Theoretically, the transport equation needs only two measurements to determine three unknowns - intensity and phase. However, three measurements is preferred since it allows the gradient measurement $dI/d\lambda$ to be centered about the central plane. Next, we calibrate the spectral channels to equivalent defocus measurements to recover QPI by solving the linearized transport equations.

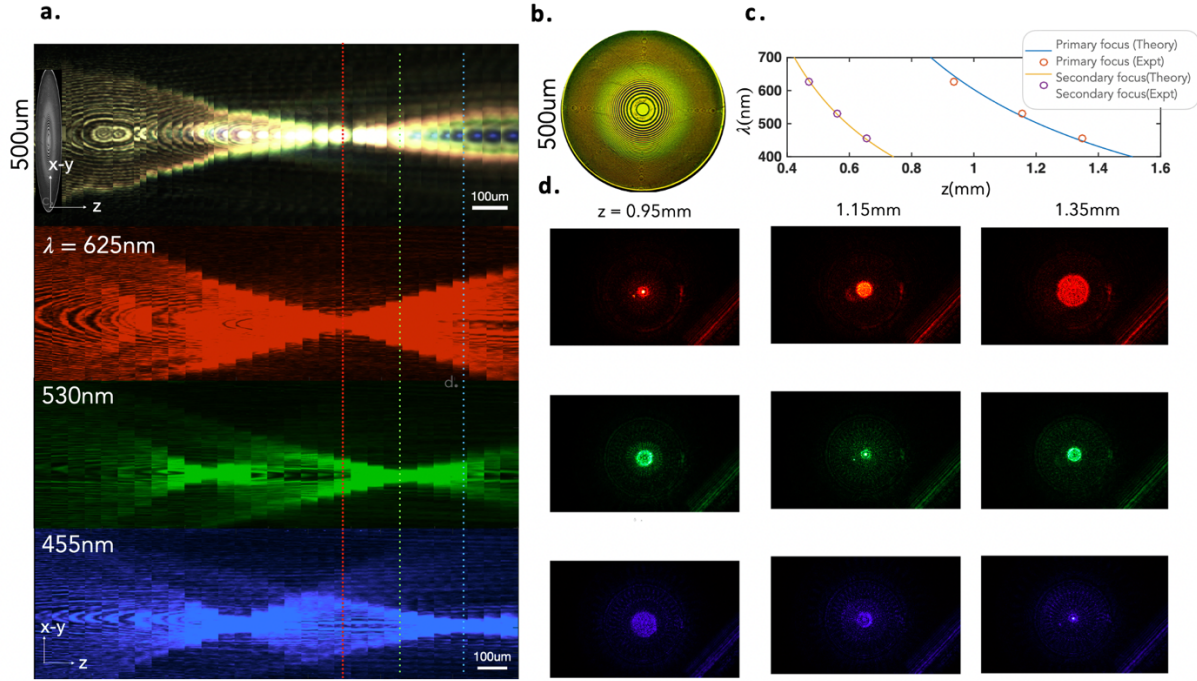

*Figure 1: Metalens Spectral Characterization a. The metalens is illuminated with collimated light from a LED source with wavelengths centered at red (625nm), green (530nm) and blue (455nm) respectively. Raw 2D intensity images are sliced column-wise and stacked to create the through-focus (x-y vs z) profile showing spectral dependence of focal length. The meta-optic primarily focuses at 1mm, with a weak secondary focus at 0.5mm for blue and green ligth.. b. Microscope image of the fabricated hyperboloid lens. c. The focal length along propagation direction as a function of wavelength $\lambda$ shows almost perfect fit to $\lambda \propto 1/z$. d. The 2D PSF as a function of the focal distances for the RGB LEDs. The red comes to focus first at 0.95mm (top row), followed by green at z= 1.15mm (middle row), and blue at z = 1.35mm.*

**Calibration of spectral vs. focal distances**

Our spectral QPI is first calibrated using a precision diffuser (see Methods) as a designer disordered media, since defocus phase contrast is maximized with a strongly scatterer. Once we've measured the dependence of spectral channels on focal distance (Fig. 2), we solve the phase using the Transport of Intensity Equation for coherent propagation as discussed previously. The recovered height (h) from the solved quantitative phase ($\phi$) is then validated with a Seimen's star target where the ground truth is known (Fig 3).

The metalens microscopy setup is shown in Fig 2a. The metalens acts as a single lens imaging system with unit magnification. Collimated light from a white LED is incident on the planar, pure-phase diffuser. The diffuser is placed two focal lengths anterior to the metalens, forming an image two focal lengths posterior to the metalens with unit magnification. The microscope objective is placed with its working distance at the plane where the image is formed by the metalens. Since the target is a phase target, it is perfectly transparent at focus, with $I(x,y) = 1$. The microscope objective relays the image formed by the metalens to a standard RGB color camera (See Methods / S2).

Positive and negative defocus give us strong focal line networks called caustics, similar to the bright focal lines in the bottom of a wavy pool. These caustics have complementary patterns on either side of focus (Fig. 2b,c,d), with junctions of lines on one side of focus located at gaps between lines on the opposite side of focus. Supplement S3 shows a continuous transformation between caustics formed with and without the metalens in place, highlighting the aberrations in the metalens as a deformation of caustic patterns.

The defocus distance at which caustics first appear is observed to be about $\pm 100 \mu m$ estimated by the Talbot distance $\delta z = 2p^2/\lambda$ [34], where p is the nominal feature size (about $7 \mu m$ for the diffuser) and $\lambda$ = 530nm as the nominal wavelength. The lateral resolution and field of view are preserved with the relay optics since the NA of the relay optics is greater than the NA of the meta-lens. The red and green channels are shown to capture complementary caustic patterns, corresponding to positive and negative defocus.

By calibrating with defocus measurements of the diffuser, we find that the red and green channels correspond to two defocus planes separated by $180 \mu m$, simultaneously. We are thus able to utilize the two color channels as a through-focus measurement to compute the quantitative phase and hence the entire electric field or light field, The digital holographic reconstruction or refocusing is performed by convolution with the defocus PSF (described in detail in a following section). Since two color channels or defocus images are sufficient to compute the phase, we rely on just the red and green channels for most of the phase retrieval performed with our metalens.

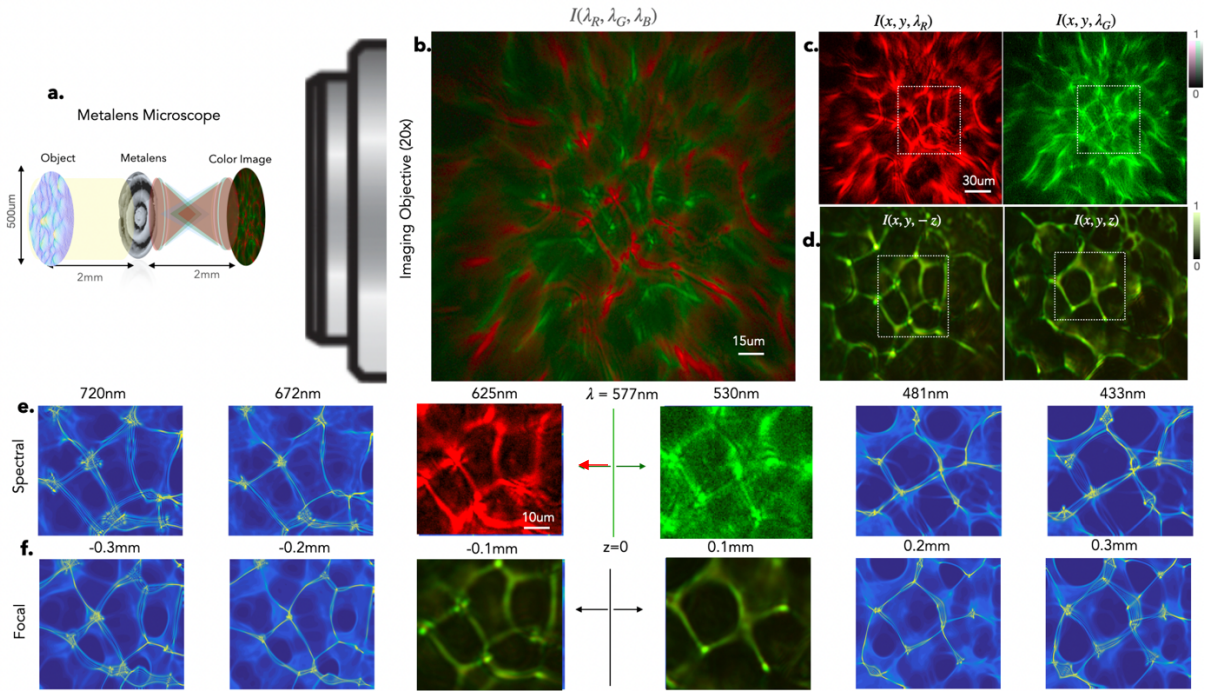

*Figure 2: Spectral Focal Equivalence* ***a.*** *The metalens microscopy setup in imaging condition with the $20\times$ microscope objective (drawn to scale). The metalens is 2f (2mm) away from the phase target, in this case a precision diffuser, with an inverted image formed at 2f from the metalens. The microscope objective magnifies the real image by $20\times$ after passing through a tube lens onto a color CMOS camera.* ***b.*** *The single shot color image captured on to the camera.* ***c.*** *Red and green simultaneously capture positive and negative defocus images on either side of the diffuser.* ***d.*** *Equivalent positive and negative defocus measurements demonstrate that a 95nm spectral shift ($\lambda_R = 625nm$, $\lambda_G = 530nm$) is equivalent to a $180\mu m$ defocus ($z = \pm 90\mu m$).* ***e.*** *Phase retrieval enables numerical propagation of the electric field to extrapolate the image formed for wavelengths separated by ~48nm over the entire visible spectrum* ***f.*** *Equivalent defocus images at 0.1mm steps since $\lambda f = constant$.*

**Quantitative Phase Microscopy**

Next, we image quantitative phase using the camera's color channels in a single shot. We use the same broadband LED source, scattered by the target and imaged through the metalens, then relayed by a microscope objective onto the Bayer filter on the camera (Fig. 3). The color camera integration time for our single shot acquisition is <100ms which enables phase imaging at video rate. We calibrate with a spectrometer to estimate the effective wavelengths to be 625nm, 530nm and 455nm (See Methods). We then use the intensity gradients vs wavelength ($dI/d\lambda$) in the TIE to compute the quantitative phase $\phi$ [33] ; subsequently we validate our solved phase against the measured ground truth (Fig 3b-d). A Seimen's star phase target (Benchmark Inc.™) is used to validate our method. The Seimen's star rotates in opposite directions on either side of focus (See S4). Additionally, once the phase has been solved, we validate that the phase gradients vs wavelengths extracts the edges of the star target (Fig 3b, bottom) as predicted by the Transport of Phase Equation $(d\phi/d\lambda \propto \hat{\nabla}\phi.\hat{\nabla}\phi)$.

The retrieved phase is shown to be accurate by two methods – first by measuring the Seimen's star target with a commercial phase imaging camera for the ground truth (See Supplement S7 for SEM images of the Seimen's star). Second, by digitally propagating the recovered electric field and comparing with measured defocused images (Figs. 3f, 3g). Our phase retrieval method shows $0.1\lambda$ sensitivity of the retrieved height, limited only by the noise in the measured intensity derivative dI/dz (Fig. 3e top). The differential method automatically discards systematic noise that is common to the intensity as it propagates through focus, yet has remnant noise due to other sources. The calculated contrast transfer function is shown in Fig. 3e (bottom), by radially averaging the QPI image to estimate the contrast of the recovered phase as a function of the grating pitch. The resolution is estimated to be $\sim 1\mu m$ at 10% of the maximum contrast, comparable to the theoretical limit of $\lambda/2NA = 1.1\mu m$. However, the contrast is sensitive to the height of the target. For our measurement, targets under 100nm height showed poor signal contrast limited by the low coherence of the LED source[35] as well as other noise sources (thermal noise, dark current, photon shot noise, ambient noise in the room etc) . Hence the practical sensitivity of the method is closer to $\sim 0.2\lambda$. However since propagation performs a high frequency filtering of the phase, smaller features can still be extracted using larger defocus measurements [36].

**Holographic Reconstruction.** Once the phase retrieval gives us the full complex electric field, we have enough information to reconstruct the 3D light-field for a coherent beam (at the nominal wavelength or defocus). We can thus digitally refocus the object in the intermediate planes to extrapolate the scattered light field in the entire volume. Hence a digital holographic reconstruction for the wavelengths in-between the (R, G, B) measurements is numerically interpolated as shown in Fig. 3g. Digital refocusing is computed as the 2D convolution of our phase object's complex value field $\widetilde{E_o}$ with the defocus PSF as $\widetilde{E_z}(x,y) = \widetilde{E_o}(x,y) * h_z(x,y)$, where $h_z(x,y) = \frac{e^{ikz}}{i\lambda z}e^{i\frac{2\pi}{\lambda}\frac{x^2+y^2}{2z}}$ is the Green's function or PSF for defocus z (assuming the propagation length is large, i.e., $z^2 >> x^2 + y^2$) [37]. We observe that the conjugacy of $\lambda$ and z is preserved since they are coupled as equivalent variable $\zeta = \lambda z$ in the propagation kernel. The spectral-focal equivalence allows validation of the measured phase by digital propagation with the defocus kernel, followed by comparison with the measured defocus image stack (Fig. 3g). The through-focus image stack is considered as the ground truth for validation: we then computationally simulate equivalent spectral shifts and validate them against equivalent defocus planes by finding identical features. Hence, we are able to compute a hyper-spectral dataset starting from the single RGB image captured by the color camera, that can test the validity of our phase retrieval. Since phase retrieval is an ill-posed problem with multiple solutions satisfying given measurements, a holographic reconstruction gives us a quantitative estimate of the accuracy of our solved light-field.

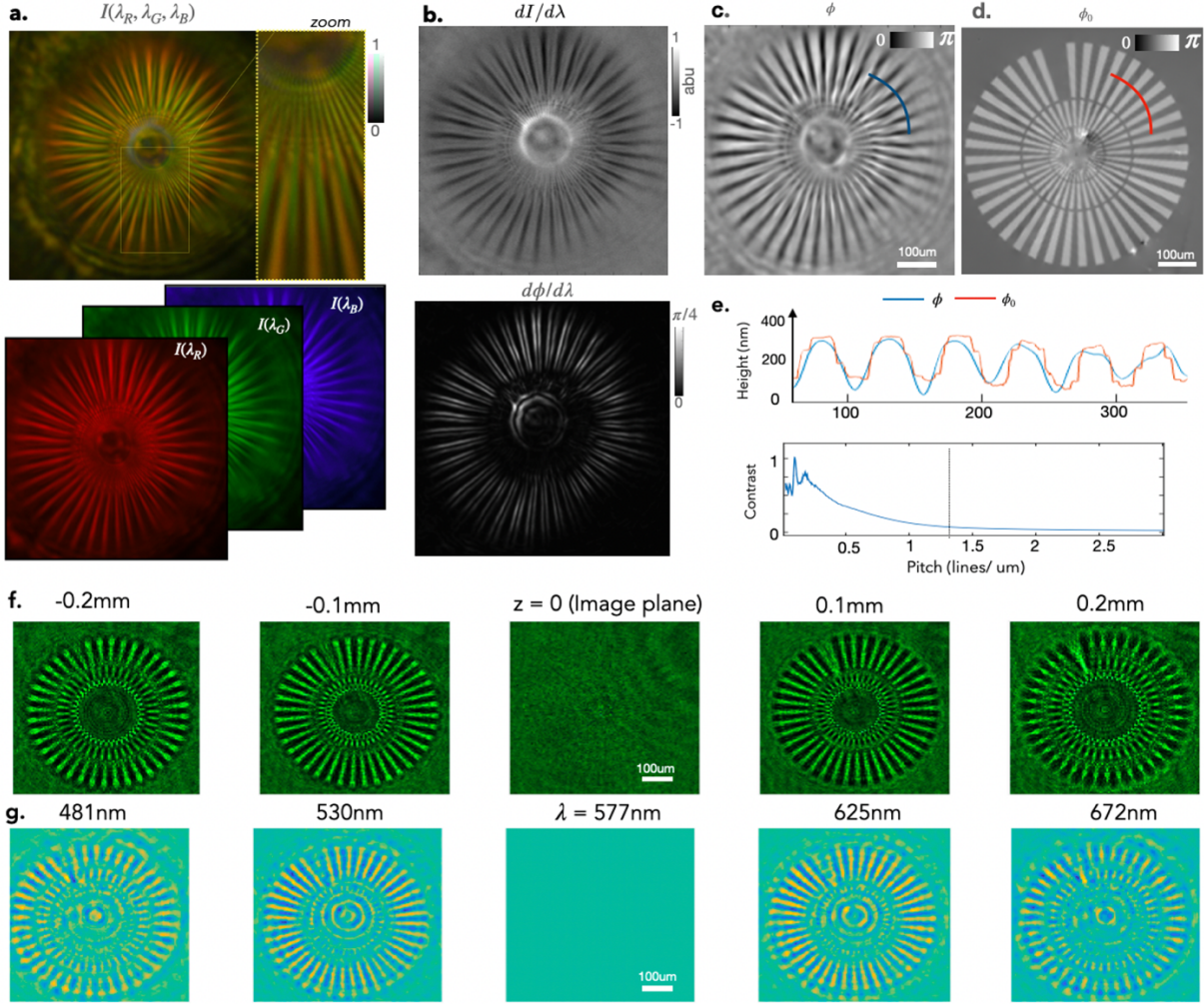

*Figure 3: **Quantitative Phase Imaging** using a metalens for a 350nm tall Seimen's star target. **a.** Color image captured by a camera (top) with an integrated color filter gives RGB images in three separate channels (bottom). **b.** The intensity difference between the color channels ($dI/d\lambda$) gives effective intensity change vs color (top), while the phase change with color $d\phi/d\lambda$ is computed post QPI (bottom) to reveal the spatial phase gradients as $d\phi/d\lambda \propto \widehat{\nabla}\phi.\widehat{\nabla}\phi$ **c.** The recovered QPI image gives exact height measurement of the test target **d.** The ground truth is measured on a Nikon microscope with a commercial phase imaging camera **e.** The recovered phase (blue) is compared with the ground truth (red) along the cutlines in c. and d. after converting to equivalent height value as $h = \frac{\phi}{\Delta n}\frac{\lambda}{2\pi}$. Plot in blue shows $0.1\lambda$ sensitivity of the QPI. The contrast is computed as a radial average of the recovered phase. **f.** Measured intensity images of 350nm tall Seimen's star target as a function of defocus for $100\mu m$ defocus steps. Pure phase targets are invisible at focus. Radial spokes arise from the center and move outward with increasing defocus. **g.** Holographic reconstruction of the spectral stack from a single color image captured by our metalens. Once QPI yields the coherent electric field, it can be propagated digitally to compute intensity at intermediate wavelengths, then compared with equivalent defocus images in **f**.*

**Quantitative Phase Endoscopy**

The metalens microscopy setup is subsequently translated to a coherent fiber bundle endoscope (Fig. 4, 5). The fiber bundle consists of 18000 single mode fiber with diameter of $3\mu m$ each in a 1mm$^2$ area. Each fiber within the bundle acts as a photodiode, retaining only the intensity of the incident light for each color, while losing the information about direction ($\widehat{\nabla}\phi$) or phase ($\phi$). However, since the R, G, B *intensity* $I(\lambda_r, \lambda_g, \lambda_b)$ channels encode phase $\phi$, we can afford to lose the coherent phase as long as the spectral channels are reliably transmitted through the fiber bundle, with the phase encoded in the relationship between the RGB intensities. Additionally, fiber spectral intensity transmission is more resilient to bending and vibration. Hence the spectral encoding of phase is more robust to real-time maneuvering than typical calibration based methods that invert a transmission matrix which has to be re-calibrated every-time the fiber moves or bends.

We apply the TIE based phase retrieval technique for phase retrieval through the coherent fiber bundle. With only a single color image, the phase is computed and validated with the ground truth (Fig. 4). For the first proof of principle, we assume transmission geometry and a point source. The resolution is limited by the individual fiber core diameter ($3\mu m$) after magnification by the metalens. We choose an imaging geometry with 5x magnification from the single metalens imaging system for calibration in Fig. 4, limiting the spatial resolution to about $6\mu m$. For measurements of a biological sample in Fig. 5, the magnification is reduced to unity for increased field of view.

We finally image a volumetric biological target to extract structural and topological information from a single white light image. A spirogyra alga (*Spirogyra porticalis*) is chosen because of its distinct spiral chloroplasts, $10\mu m$ width resolvable by the endoscope, and three dimensional winding geometry. Fig. 5 shows recovered phase and height maps for the spirogyra, showing structural features such as strands in different planes and topology of the spiral chloroplasts. With the endoscope added, although we lose resolution, we are still able to identify the elliptical nuclei (zygospores) and distinguish top and bottom strands at a crossing. Coherent fiber bundles typically suffer from honeycomb artifacts due to the arrangement of the fibers in the bundle. Our spectral-focal equivalence based single shot phase retrieval automatically filters out the shared artifacts and common noise in the color channels due to its difference based computation. Additionally, the regularization in the solver effectively low pass filters out fiber packing geometry in the recovered phase while retaining the quantitative nature due to calibration with a known target . The regularization parameter is kept fixed between the calibration and validation datasets, assuming all other experimental conditions to be constant. The phase sensitivity is within the lower and upper bounds dictated by noise and dynamic range of the camera respectively. Additional frequencies can be recovered in a hyper-spectral dataset with multiple color channels at the camera or with a swept spectrum source.

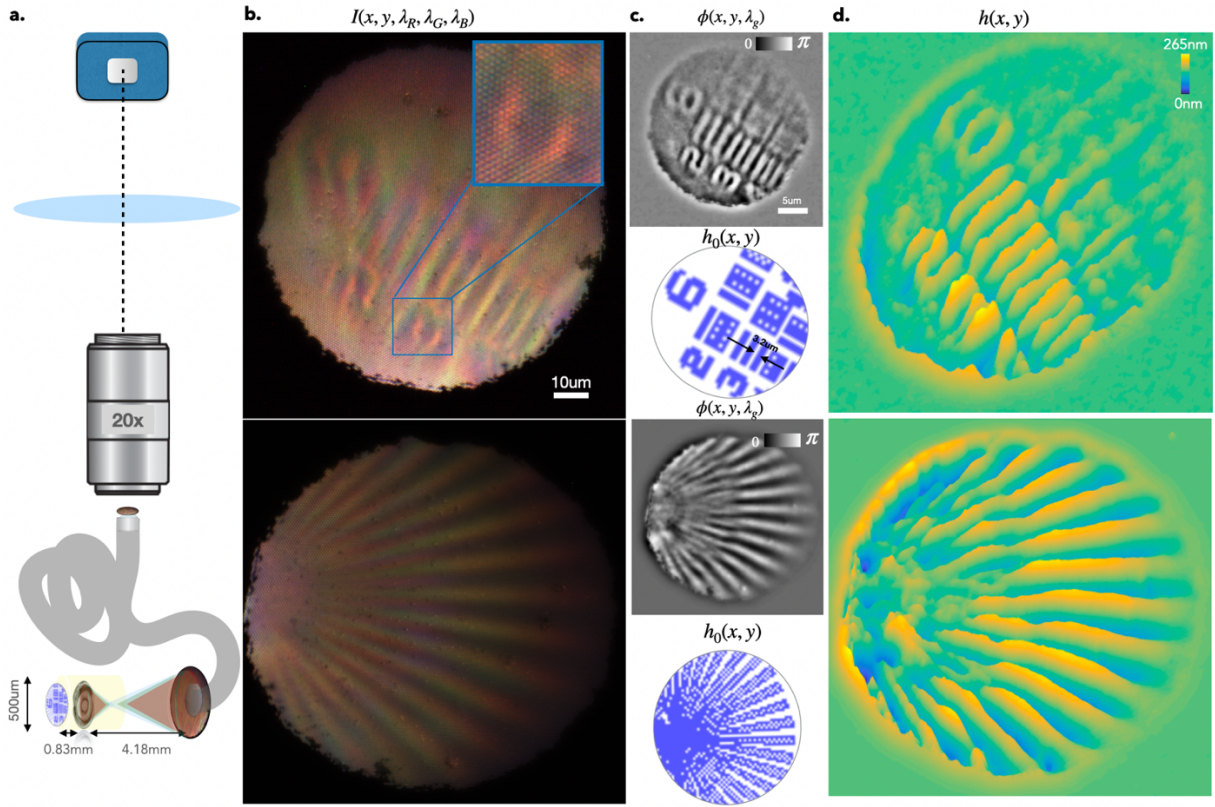

***Figure 4. QPI through CFB.*** *Validation of quantitative phase imaging through a coherent fiber bundle. The 3um individual fiber cores preserve spectral intensity but scramble spatial phase.* ***a.*** *The metalens magnifies the sample/target by 5x onto the face of a coherent fiber bundle. The fiber bundle acts as a photodetector array at the distal end (sample and metalens) and as a ensemble point source emiter at the proximal end (camera). The image transmitted by the fiber bundle is relayed by a microscope objective to the color camera, with a further magnification of 20x.* ***b.*** *The distal end of the fiber bundle preserves the intensity information for each color but scrambles the phase. Using the correlation between the color channels we are able to extract the quantitative phase of the sample before passing through the fiber.* ***c.*** *Quantitative phase at the distal end of the fiber from measurements captured at the proximal end (top); ground truth as measured by manufacturer Benchmark Technologies Inc. (bottom)* ***d.*** *The height map is extracted as $\phi/\Delta n/k_0$ at the nominal wavelength. Scale bars are at the sample plane w/o magnification.*

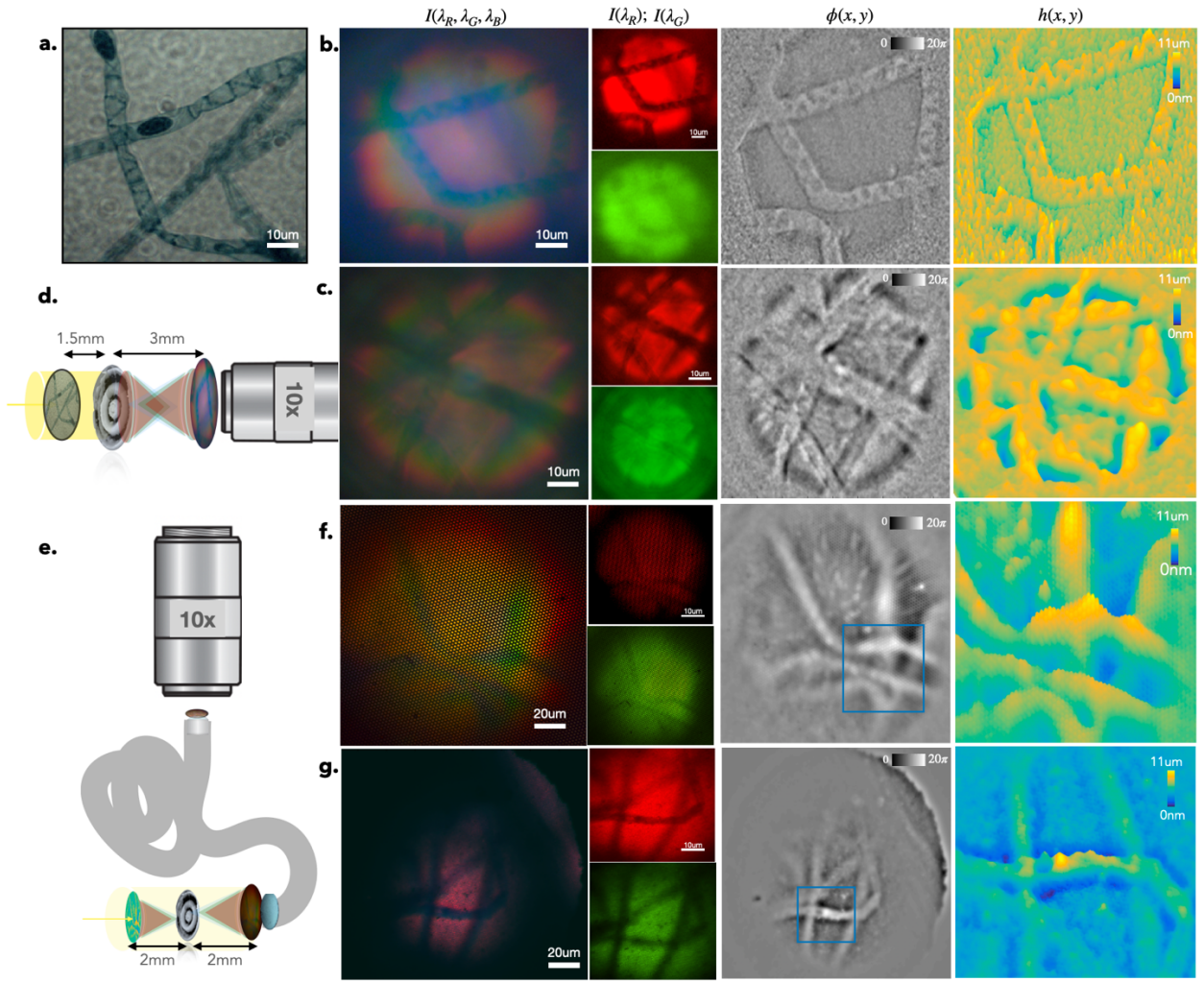

*Figure 5. QPI for a Spirogyra sample. **a.** Ground truth of the spirogyra alga shows spiral chloroplasts and dark beadlike zygoplasts. **b,c,f,g.** Captured White light image $I(\lambda_R, \lambda_G, \lambda_B)$, two color channels $I(\lambda_R); I(\lambda_G)$, recovered quantitative phase $\phi(x,y)$, and height map $h(x,y)$. **b.** Metalens based single shot QPI shows structural information, such as individual chloroplast strands. **c.** Spirogyra on either side of the focal plane show opposite phase curvature, seen as depressed or elevated grooves. **d.** Imaging configuration for metalens only imaging system for b, c. **e.** Imaging configuration with metalens and coherent fiber endoscope. **f.** QPI through an endoscope is able to distinguish top and bottom spirogyra in a crossing **g.** Structure of individual zygoplasts visible in the phase image even in low brightness conditions.*

**Discussion**

We have reported a single shot technique for recovering quantitative phase for suture-less endoscopy (<1mm incision) with a single metalens in transmission geometry. We utilize the chromatic dependence of focal distance inherent in a hyperboloid metalens to encode three defocus planes into RGB color channels measured by a color camera; the through-focus intensity enables computational recovery of quantattive phase. The chromatic aberration in the metalens is tailored to have an exact inverse linear dependence with the three wavelengths of interest. The three wavelengths are determined by the camera's color filters (Bayer sensor) and the source spectrum. Our method is shown to be applicable with low cost white light LEDs and any off-the-shelf color camera, enabling potential translation to mobile Point of Care Diagnostics [38]. Our phase sensitivity and field of view with only a single optical element holds potential for commercial holographic endoscopy while capturing the entire 3D light-field even with low coherence or high dispersion. The phase contrast mechanism is sensitive to a tenth of wavelength in the presence of noise at room temperature, with nominal (<100ms) camera exposure time to support video rate acquisition. The light budget was found sufficient with the tabletop LED source to have enough Signal to Noise ratio for quantitative estimates of the phase at $\lambda/5$ height of the calibration targets. The lateral resolution was seen to be diffraction limited by the metalens aperture at $\lambda/2NA = 1\mu m$ for the microscopy configuration, and by the fiber core diameter (*3um*) in the endoscopic configuration with a fiber bundle. Hence our QPI enabling metalens can be added onto any endoscopic modality while maintaining the native resolution. Several challenges still need to be addressed for translational applications of our method. The single lens system used for phase imaging is simple in terms of alignment and complexity, yet suffers from non-telecentricity that has to be compensated computationally[39]. In a single lens system, the image is formed on a spherical wavefront – the magnification varies as a function of the field of view (lateral) as well as defocus (transverse). Hence the magnification varies slightly between the red, green, and blue channels, causing a spherical aberration in the phase. A potential solution would be to use a grin lens at the distal end of the endoscope that collimates the wavefront from the metalens to form a telecentric imaging system.

Neverthelss, we have shown for the first time to our knowledge a single shot spatial mapping of relative phase shifts in white light transmitted through a specimen using a metalens at distal tip of a CFB and a color camera as the proximal end of CFB. Our method is implemented in single-pass transmission geometry, even though most in-vivo applications require reflection mode imaging. Quantitative phase images lose contrast in double pass reflection geometries due to multiple scattering in the sample; however modalities that utilize backscattering from deeper layers can still enable phase imaging in single pass through superficial layers before being collected by the distal metalens tip. Backscattering can be achieved by various means – by inducing auto fluorescence in deep tissue[40], by creating an oblique offset between the illumination and collection pathways [41], or by using deeper diffuse reflective layers as a backlight (for example in ophthalmoscopy where light from back of the retina acts as a diffuse reflective screen to illuminate the retina for single pass QPI [42]).

## Materials and Methods

### Solving the Transport of Intensity Equation

To solve the transport of intensity equation, we substitute the Poynting vector by the gradient of a scalar potential $I\widehat{\nabla}\phi = \widehat{\nabla}\psi$ to arrive at a Poisson equation of the form $dI/d\lambda = \nabla^2\psi$. First, the Laplacian $\nabla^2$ is integrated in the Fourier domain. The double derivative in the two-dimensional Laplacian $\nabla^2 = \nabla_x^2 + \nabla_y^2$ Fourier transforms as a parabola ($k_x^2 + k_y^2$), that is divided out as a frequency domain deconvolution or integration (with a constant regularizer added to prevent division by zero) to obtain $\psi$. Subsequently $\phi$ is solved from $\psi$ by solving another Laplacian $\nabla^2\phi = \widehat{\nabla}.\left(\frac{\widehat{\nabla}\psi}{I}\right)$. An iterative procedure that updates a residue in the intensity gradient after forward propagation of the solved fields through the TIE and TPE is applied repeatedly until convergence. The residue is solved again with the TIE solver to give a corrected phase estimate. The iterative solution helps remove artifacts in the presence of absorption, recovers vorticity, and improves fidelity of the quantitative phase [33].

### Metalens Fabrication

We fabricated the meta-optical holograms using a 500-micron thick fused silica wafer and depositing 600 nm of Silicon Nitride using Plasma enhanced chemical vapor deposition at 350-degree centigrade. A 300 nm thick layer of ZEP 520A followed by a thin film of anti-charging polymer (DisCharge $H_2O$) was spun coat on top of $Si_3N_4$ thin film. Next, the lens patterns are written by electron beam lithography (JEOL 6300) at a beam energy of 100 keV, beam current of 8000 pA, and a base dose of 275 µC/$cm^2$ and appropriate proximity effect corrections. The resulting designs are developed in a solution of amyl acetate and Iso-Propyl alcohol. The exposed and developed samples are then placed in a physical evaporator to deposit roughly 48 nm of aluminum oxide by electron beam evaporation. The dissolution of the remaining resist performs lift-off in N-Methyl-2-pyrrolidone (NMP) at 90°C for 12hours. Finally, the pattern is transferred from the aluminum oxide mask to the $Si_3N_4$ by using a fluorine based RIE process (Oxford) leaving a total thickness of 10 nm of Alumina over 532 nm of $Si_3N_4$.

**Metalens Characterization** is performed with RGB fiber coupled LEDs from Thorlabs. Red M625F2, Green M530F2, Blue M455F3 LEDs are coupled to a 100um single mode fiber core for ensuring coherence, since we are performing coherent phase retrieval. The collimating lens is chosen to be 2″ diameter, 30mm focal length achromat. For enhanced signal to noise ratio in the phase retrieval, higher coherence is preferred at the cost of more speckle in the image. A smaller diameter fiber and longer focal length collimating lens would yield sharper phase images, since incoherence effectively leads to image blurring. However higher coherence leads to lower photon count, hence requiring larger camera integration time. The model of the camera used is Allied Vision GT1350.

**Imaging Setup**

Experiment performed with collimated Light emitted from the Thorlabs white light fiber coupled LED (MCWHF2) incident on the sample, followed by imaging by the meta-lens. The LED is operated at nominal power (~27mW) spread over a 1 inch collimated beam for an operating power concentration of ~42W/m$^2$, within acceptable range for most tissue is the body for short periods.

Alignment with white light can be performed by making sure that the focal spot for one color shows three co-centric circles for the other two colors. The spectral-focal calibration is performed with a Newport 10DKIT-C1 $5°$ precision diffuser characterized by a $5°$ beam waist of the diffused beam for plane wave incidence, The phase calibration sample is manufactured by Benchmark Technologies Inc. The image plane is relayed by the 20x objective to the camera. The ground truth phase measurement for the calibration target is performed with the Phasics© SID4 HR phase imaging camera at the output port of a regular transmission microscope.

The metalens imaging setup includes an infinity-corrected 20x objective (Nikon Plan Fluor 20x, 0.50 NA), which collects light scattered by the metalens at a distance of 2mm. The objective and tube lens assembly enables a 20x magnification of the image formed by the metalens. The tube lens is from Thorlabs (AC254-200-A-ML) with focal length f = 200 mm. The sensor is 3.2MP Infinity5-3 CMOS color camera with pixel size of $3.2\mu m$. Frame rates for all the measurements are >10Hz, enabling real time phase imaging in an in-vivo setting.

For imaging with an optical fiber (Schott RLIB CVET, 1.05 × 910, 7.6 M, 18K19, QA.90 ) separating the metalens and the camera, both ends of the fiber are placed on a 3D motion stage for alignment; the sample and the metalens are also on motion stages. An alternative strategy with fewer moving parts uses a 2mm thick O-ring to separate the metalens and the sample, which are then mounted with scotch tape in a fixed imaging configuration (See Supplement S2 for full schematic). Spirogyra samples were purchased from Carolina biological supply (Item# 296548). Two slides are stacked face to face with the Spirogyra sandwiched between the glass to accentuate three dimensional structures. The distal end of the fiber looks at the sample through the metalens; the proximal end of the coherent fiber bundle is directly observed under a Nikon Eclipse LV100 microscope with the objective dependent magnification as indicated in the figures.

**Estimating effective wavelengths for phase retrieval**

The quantitative phase retrieval needs accurate values of the effective wavelengths $\lambda_r$, $\lambda_g$, $\lambda_b$ for the RGB intensity channels of the color camera, respectively. Since we use a broadband white light LED source, the camera and source spectra have to be multiplied to find the peak effective wavelengths corresponding to Red, Green and Blue. We use a spectrometer to measure the source spectral response, which is multiplied by the standard response curve of the camera's color filter to find our peak wavelengths to be reasonably close to the wavelengths used for metalens characterization i.e. 625nm, 530nm and 455nm respectively.

**Acknowledgements**

The research is supported by NSF- 2040527. Part of this work was conducted at the Washington Nanofabrication Facility / Molecular Analysis Facility, a National Nanotechnology Coordinated Infrastructure (NNCI) site at the University of Washington with partial support from the National Science Foundation via awards NNCI-1542101 and NNCI-2025489.

We would like to thank Fiona Xi Xu and Dr. Dan Fu for access to their Phasics phase imaging system for ground truth validation of the phase calibration target.

Supplementary information accompanies the manuscript on the *Light: Science and Applications* website (http://www.nature.com/lsa)

**Supplementary Information**

**Quantitative Phase Imaging with a Metalens**

Aamod Shanker[1], Johannes Froech[1], Saswata Mukherjee[1], Maksym Zhelyeznyakov[1],

Eric Seibel[3], Arka Majumdar[1,2*]

[1]Department of Electrical and Computer Engineering, University of Washington, Seattle, WA
[2]Department of Physics, University of Washington, Seattle, WA
[3]Department of Mechanical Engineering, University of Washington, Seattle, WA 98105
* Corresponding author: arka@uw.edu

**S1 : Derivation of the Transport of Intensity and Phase from Maxwell's equations.**

We will derive the transport equations for free space optical propagation under the paraxial approximation. We start from electromagnetic waves in free space described by Maxwell's equations. The Helmholtz Equation is derived by time averaging the Maxwell's equation to get its temporally stationary form. The Helmholtz equation describes the time averaged distribution of the electric field in three dimensional space as the elliptical partial differential equation,

$$(\nabla_{xyz}^2 + k^2)\vec{E} = 0 \quad \text{------ (1)}$$

where $\vec{E}$ is shorthand for $\vec{E}(x,y,z)$, the electric field vector distribution in space, $k$ is the spatial momentum or wave-vector with x, y, z components given by $k_x, k_y, k_z$ respectively, with $|k| = 2\pi/\lambda$. Additionally $k^2 = k_x^2 + k_y^2 + k_z^2$ and $\nabla_{xyz}^2 = \nabla_x^2 + \nabla_y^2 + \nabla_z^2$ is the 3D scalar Laplacian. Also $\hat{\nabla} = \frac{\partial}{\partial x}\hat{x} + \frac{\partial}{\partial y}\nabla\hat{y}$ is the 2D in-plane gradient vector and $\nabla^2 = \nabla_{xy}^2 = \nabla_x^2 + \nabla_y^2$ is the 2D scalar Laplacian. Under the paraxial approximation $k_z \gg k_x^2 + k_y^2$ we assume that the wave is mainly propagating along a single direction z. Hence the electric field can be approximated as $\vec{E}(x,y,z) = U_0(x,y,z)e^{ikz}(\hat{x} + \hat{y})$, with the electric field polarized normal to propagation in z.

Substituting into the equation (1) and expanding the z derivative by the chain rule,

$$(\nabla_{xy}^2 + \nabla_z^2 + k^2)U_0(x,y,z)e^{ikz} = 0$$

$$\Rightarrow [\nabla_{xy}^2 U_0 - k^2 U_0 + 2ik\frac{dU_0}{dz} + ik\frac{d^2U_0}{dz^2} + k^2 U_0]e^{ikz} = 0$$

$$\Rightarrow [\nabla_{xy}^2 U_0 + 2ik\frac{dU_0}{dz} + ik\frac{d^2U_0}{dz^2}] = 0$$

Further making the assumption that due to the paraxial approximation is slowly varying in the propagation direction $\frac{d^2 U_0}{dz^2} \to 0$

$$\Rightarrow \nabla_{xy}^2 U_0(x,y,z) + 2ik\frac{dU_0}{dz} = 0 \text{——— (2)}$$

which is called the paraxial form of the Helmholtz equation and describes the evolution of the wave along the propagation direction z.

Now since $U_0(x,y,z)$ is a complex valued scalar field, it can be represented as its amplitude and phase components, $U_0(x,y,z) = A(x,y,z)e^{i\phi(x,y,z)}$ ——- (3)

Substitute the complex field (3) into the paraxial Helmholtz equation (2), we obtain

$$\nabla_{xy}^2 A e^{i\phi} + 2ik\frac{dAe^{i\phi}}{dz} = 0$$

$$\Rightarrow [\nabla^2 A - (\hat{\nabla}\phi)^2 A - 2k\frac{d\phi}{dz}A] + i[A\nabla^2\phi + 2k\frac{dA}{dz} + 2\hat{\nabla}A.\hat{\nabla}\phi] = 0 \text{ —- (4)}$$

Both the real and imaginary parts of Eqn (4) must be zero;

$$\Rightarrow [\nabla^2 A - (\hat{\nabla}\phi)^2 A - 2k\frac{d\phi}{dz}A] = 0 \text{ ——— (5)} \quad \text{is the \textbf{real part of (4)}}$$

$$\Rightarrow i[A\nabla^2\phi + 2k\frac{dA}{dz} + 2\hat{\nabla}A.\hat{\nabla}\phi] = 0 \text{ ——— (6) is the \textbf{imaginary part of (4)}}$$

The imaginary part of the Helmholtz equation or Eqn (6), can be reduced to the *Transport of Intensity* equation by multiplying by $A(x,y,z)$ on both sides,

$$[A^2\nabla^2\phi + 2kA\frac{dA}{dz} + 2A\hat{\nabla}A.\hat{\nabla}\phi] = 0$$

$$\Rightarrow [A^2\nabla^2\phi + k\frac{dA^2}{dz} + \nabla A^2\hat{\nabla}\phi] = 0$$

Since the square of the amplitude is the photon intensity; $A^2(x,y,z) = I(x,y,z)$ we get the

**Transport of Intensity Equation**

$$\Rightarrow [I\nabla^2\phi + k\frac{dI}{dz} + \hat{\nabla}I.\hat{\nabla}\phi] = 0$$

$$\Rightarrow \hat{\nabla}_{xy}.I\hat{\nabla}_{xy}\phi = -k\frac{dI}{dz} \quad\quad \text{——— (7a)}$$

Eqn. (7a) is the Transport of Intensity equation that describes the evolution of intensity longitudinally along z with the transverse (xy) phase and intensity gradients. The Transport of intensity if analogous to the **Continuity Equation in fluid dynamics.**

$$\hat{\nabla}.(\rho v) = \frac{d\rho}{dt} \quad \text{———(7b)}$$

where we substitute $I \rightarrow \rho$, $\hat{\nabla}\phi/k \rightarrow v$ and $dz \rightarrow dt$ to arrive at the equivalence of (7a) and (7b).

Finally, to describe the transport of momentum or phase gradients ($\hat{\nabla}\phi$) analogous to the Navier Stokes equation for fluids, we multiple the real part of the Helmholtz equation, Eqn (5) by $A(x, y, z)$,

$$\Rightarrow [A\nabla^2 A - (\hat{\nabla}\phi)^2 A^2 = 2k\frac{d\phi}{dz}A^2]$$

$$\Rightarrow [\frac{\nabla^2 A}{A} - (\hat{\nabla}\phi)^2 = 2k\frac{d\phi}{dz}]$$

substitute $A = \sqrt{I}$ and expand the derivatives in terms of $I$ and $\phi$ to obtain the

**Transport of Phase equation**

$$\frac{\nabla^2 I}{I} - \frac{(\hat{\nabla}I)^2}{2I^2} - (\hat{\nabla}\phi)^2 = 2k\frac{d\phi}{dz} \quad \text{———(8a)}$$

The Transport of Phase Equation (TPE) for light is analogous to the Navier-Stokes Equation for non-viscous fluids (Euler Equation) since they both describe to velocity ($\vec{v}$) or momentum($\hat{\nabla}\phi$) evolution.

To draw the comparison between fluid mechanics and electromagnetics, apply a spatial derivative $\hat{\nabla}(Eqn. 8a)$ to obtain

$$\frac{\hat{\nabla}\nabla^2 I}{I} - 2\frac{\nabla^2 I}{I^2}\hat{\nabla}I - \frac{2\hat{\nabla}I\nabla^2 I}{2I^2} + \frac{(\hat{\nabla}I)^2}{I^3}\hat{\nabla}I - 2(\hat{\nabla}\phi.\hat{\nabla})(\hat{\nabla}\phi) = 2k\frac{d\hat{\nabla}\phi}{dz}$$

Dropping all third order derivatives by assuming slow changes in intensity; $\nabla^n I \rightarrow 0$ where $n > 2$

$$-\frac{\nabla^2 I}{I^2}\hat{\nabla}I + \frac{(\hat{\nabla}I)^2}{I^3}\hat{\nabla}I - (\hat{\nabla}\phi.\hat{\nabla})(\hat{\nabla}\phi) = k\frac{d\hat{\nabla}\phi}{dz} \quad \text{———(8b)}$$

**Navier-Stokes Equation for non-viscous compressible fluids (Euler Equation)** is given by

$$-\frac{\hat{\nabla}p}{\rho} - (\vec{v}.\hat{\nabla})\vec{v} = d\vec{v}/dt \quad \text{———(8c)}$$

where $p$ is the internal pressure, $\rho$ is the fluid density and $\vec{v}$ is the velocity.

substituting $p \to \frac{1}{k^2}\frac{\hat{\nabla}I}{I}$, $\rho \to I$ and $\vec{v} \to \hat{\nabla}\phi/k$, the modified Euler equation(Eqn. **8c)** becomes

$$-\frac{\nabla^2 I}{I^2}\hat{\nabla}I + \frac{(\hat{\nabla}I)^2}{I^3}\hat{\nabla}I - (\hat{\nabla}\phi.\hat{\nabla})\hat{\nabla}\phi = k\frac{d\hat{\nabla}\phi}{dt} \quad .\text{———(8d)}$$

Comparing (8b) and (8d), we observe the isomorphism between the paraxial transport equation for optical phase and the Navier Stokes' equation for fluid flow.

The equivalence between light and fluid transport equations is arrived at by substituting the pressure term $p \to \frac{\hat{\nabla}I}{I}$ i.e. the pressure exerted by photons is proportional to the derivative of intensity, fairly well known in optical trapping literature. However the reciprocal correlation with intensity is less commonly known..

Additionally, since we are measuring light at steady state conditions, the dynamic variable is defocus (z) instead of time (t). Also $(\hat{\nabla}\phi)^2$ is the divergence term, that relates to the rate of outward expansion along velocity vectors. Hence light can be mathematically modeled as non-viscous fluid that exerts pressure along intensity gradients and has flow along its phase gradients.

## S2: Full Measurement Setup

The measurement was performed directly on a Nikon microscope for the collection, with an external LED illumination built next to the microscope. The metalens to sample distance was adjusted in steps of 1mm by using a one-inch lens holder O-ring (thickness of 1mm) as a spacer. The fiber was mounted on the microscope stage using a LCM-2 self centering lens mount, that can hold the fiber tip securely.

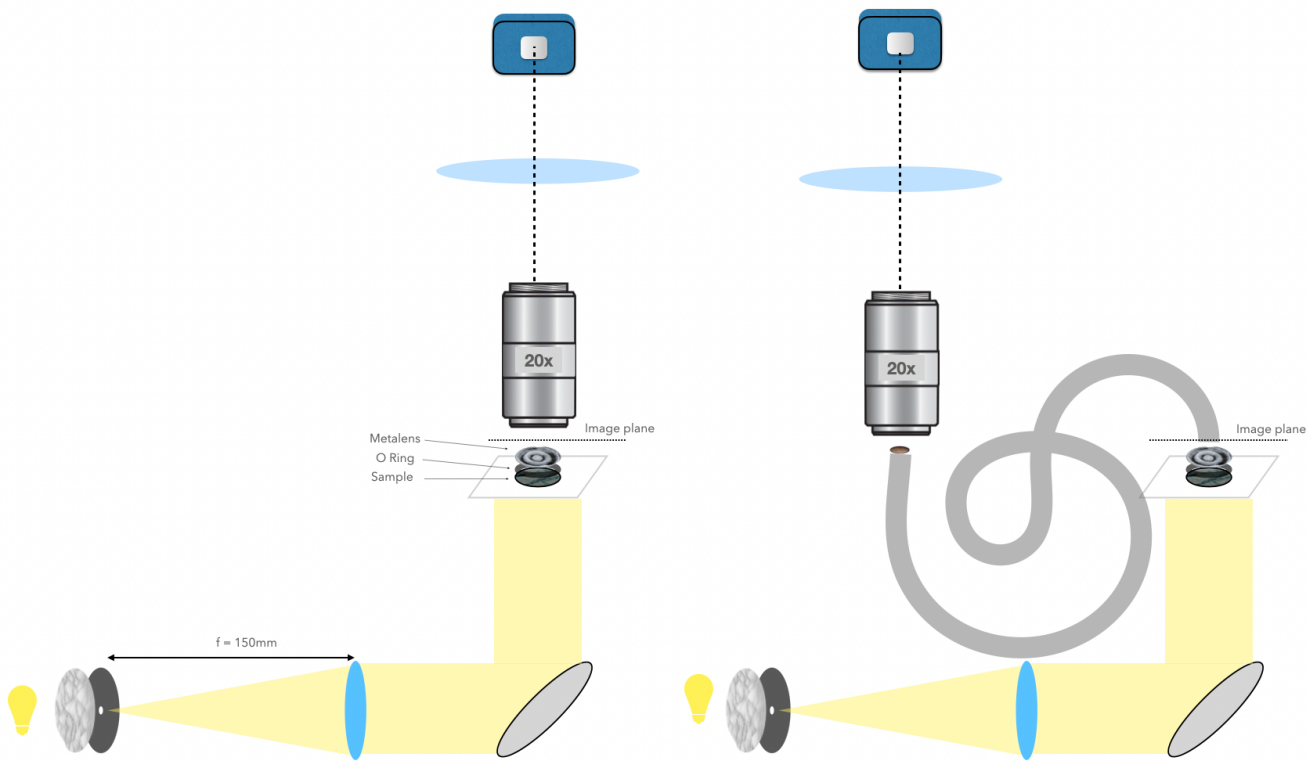

**Figure S2 Full measurement setup**

**a. Metalens Microscopy** An incandescent lamp or LED is used as a broadband source, followed by a ground glass diffuser and a pinhole aperture to ensure a homogenous beam. A collimating lens is placed at 150mm, followed by a mirror that illuminates the sample from underneath. The metalens is placed two focal lengths from the sample at the image plane, forming an image at the working distance of the objective. The objective and 150mm tube lens magnify the image formed by the metalens onto the camera by 20x (objective dependent). In certain configurations, an O-Ring is used to mount the metalens onto the sample with scotch tape at a distance of 1mm or 2mm.

**b. Metalens Endoscopy** The image formed by the metalens is relayed by the coherent fiber bundle to the microscope objective, which forms an image of the proximal tip of the fiber bundle at the camera. The resolution is limited by the size of each coherent fiber in the bundle, about 10um.

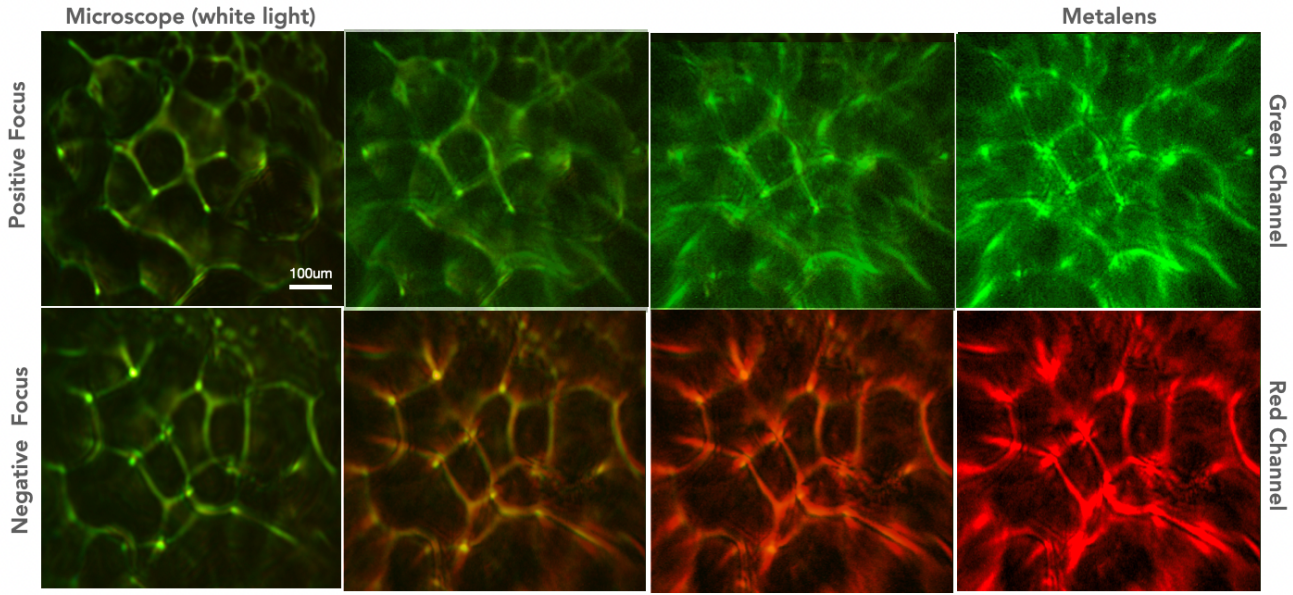

**Figure S3. Comparison of caustics: Microscope Defoii vs Metalens Color Channels.** The top row is the positive defocus in the microscope (left) , or the green channel in the metalens (right). The bottom row corresponds to negative defocus in the microscope (left), or red color channel in the metalens image (right). The columns in between the left and right show a continuous transformation starting from the ground truth microscope images into the aberrated images formed by the metalens. The junction of at-least three caustic lines can be

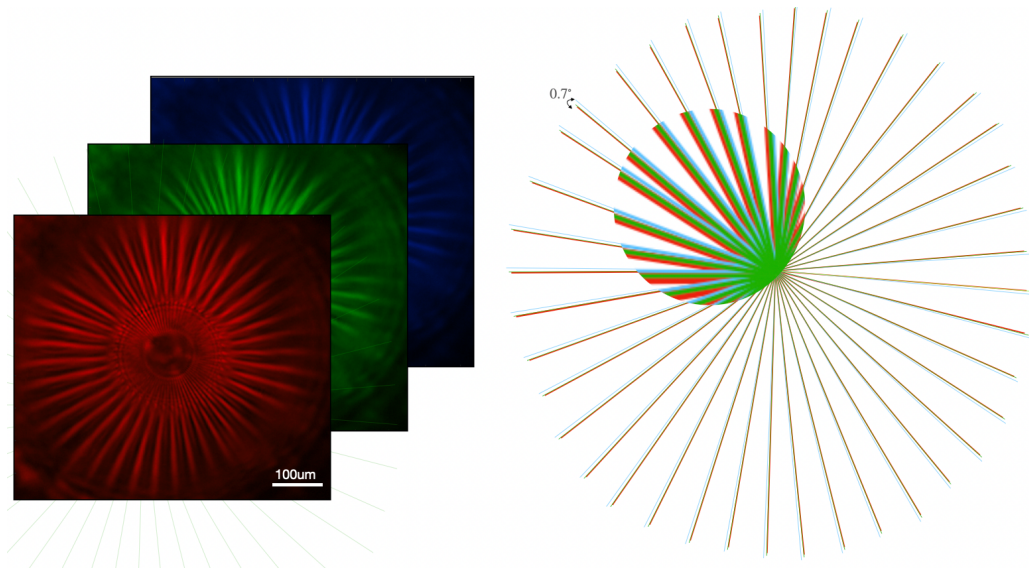

**Figure S4. Spoke locations vs Spectrum for Seimen's Star target :** The outermost spokes in Seimen's star are represented as colored lines to demonstrate the rotation of the Seimen's star target with RGB color/wavelength, identical to it's diffraction behavior through-focus. A manual fitting shows an angular rotation of 0.7° between the green and blue channels. The center is magnified to show detail.

## S5 Design of the meta-optic / SEM of fabricated metalens

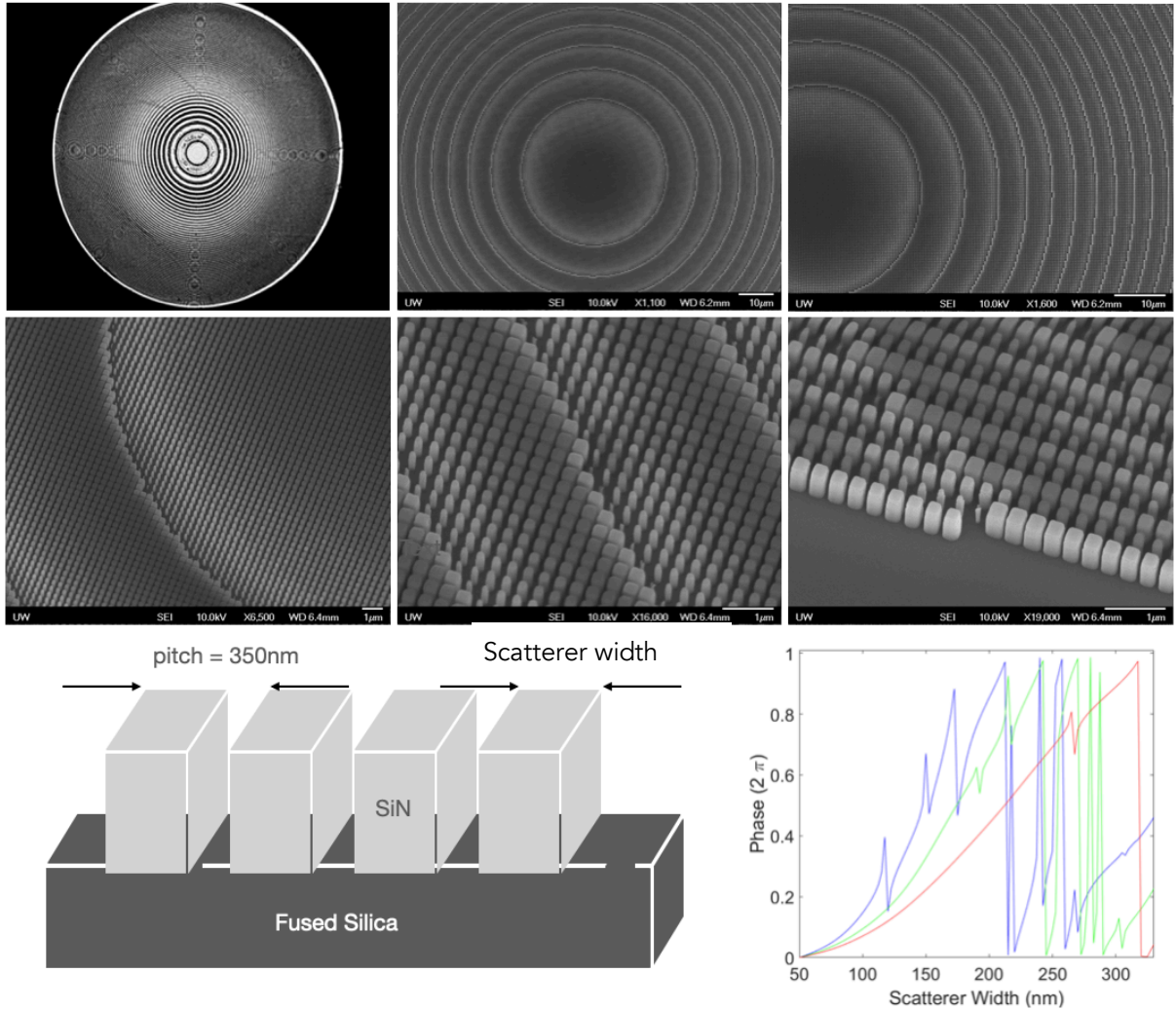

**Figure S5. Top two rows** Scanning Electron Microscopy of Fabricated Metalens shows the nano-pillars at various resolutions; scale bars at bottom right. **Bottom row** MetaOptic Design estimates the phase shift for each color (blue, green and red correspond to 455nm, 530nm and 625nm wavelengths respectively). The simulation assumes square pillars and S polarized light (invariant for P polarization under normal incidence due to the square geometry). Pillars are chosen to impart desired phase shift according to the hyperboloid phase profile at the green wavelength, taking care to avoid regions with strong fluctuation .

The computational design of the metaoptics relies on Rigorous Couples Wave Analysis (RCWA) simulations of the phase shift and absorption of periodic arrays of square nano-pillars. The simulations are performed with a periodic boundary condition assuming an infinite array of give sized pillar . Hence cross-talk between different sized pillars is assumed negligible, an assumption valid for a quasi periodic distribution where adjacent pillars have similar dimensions. With a fixed periodicity and pillar height, each meta-atom is simulated for pillar width that provides the desired phase shift. After fabrication, most of the pillars are truly closer to cylinders than squares. (Fig S1).

The metalens focal distance is  subsequently characterized with respect to the RGB wavelength

| $\lambda$(nm) | *f (mm)* | $\lambda f$ (nm-mm) |
|---|---|---|
| 625 | 0.970 | 606 |
| 530 | 1.155 | 609 |
| 455 | 1.347 | 609 |

**Table 1.** Hyperboloid lens : measured focal length dispersion shows that for our hyperboloid metalens the product of focal length and wavelength $\lambda f$ is  within 0.49% for three visible color channels.

## S6 : Single Shot phase retrieval with a binary amplitude target

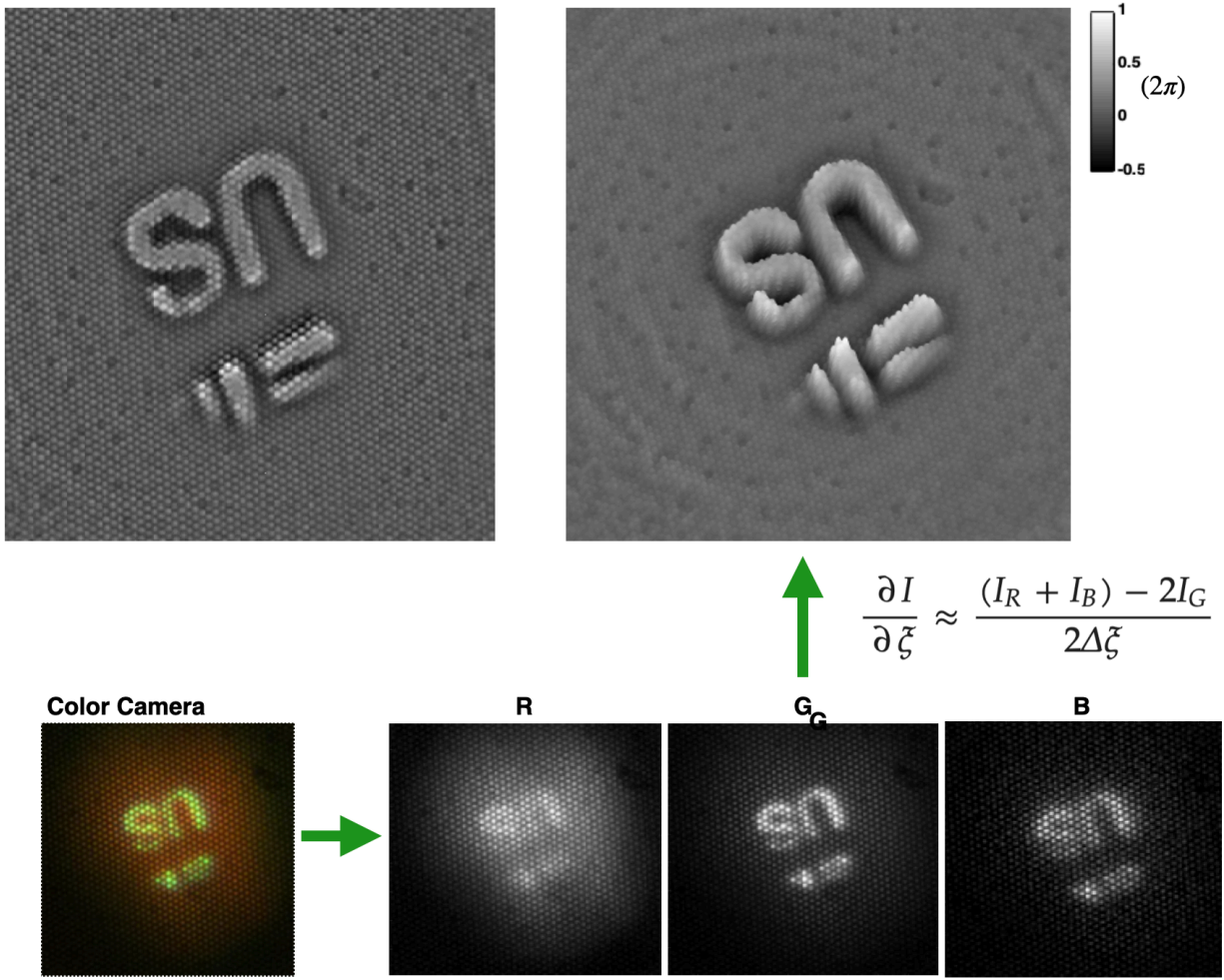

**Figure S6.** The first proof of concept measurement of phase retrieval from RGB spectral channels was performed using a binary amplitude USAF target. The bottom row shows the color image and the corresponding RGB channels of the measurement. The top row shows the retrieved phase in radian performed in the same manner as for the phase objects in the main text. $\xi = \lambda z$ represents the combined independent spectral-focal variable that encodes phase contrast in its derivative with intensity.

**S7 Scanning Electron Microscope images of Benchmark Phase Calibration Target**

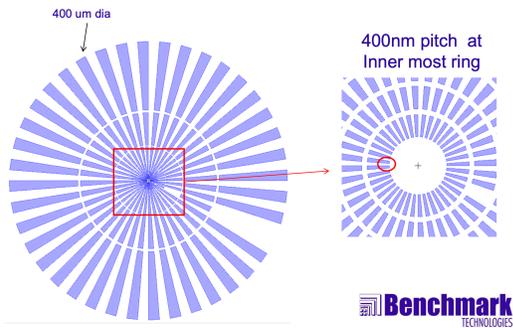
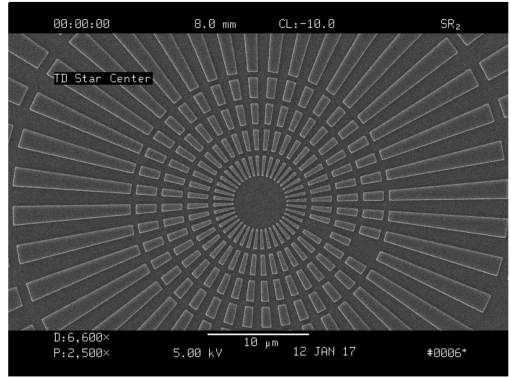

250nm nominal

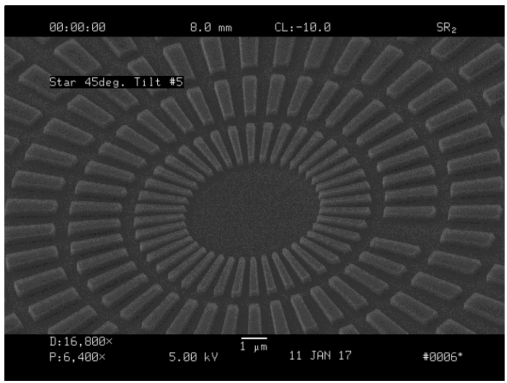
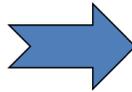
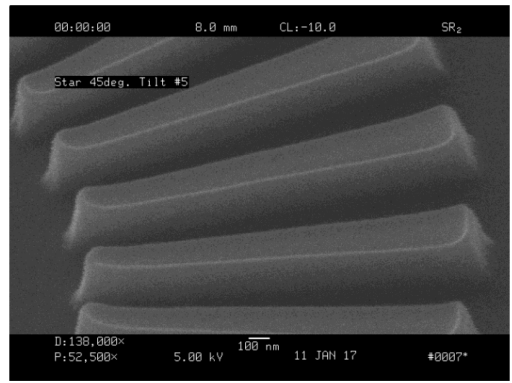

The 250nm tall phase calibration target is manufactured by Benchmark Technologies Inc. (SEM images reproduced with permission from the manufacturer).

**S8 Pseudo Code (Matlab):**

**Fresnel Propagator for digital refocusing:**

```matlab
function [Ef,x,y,Fx,Fy,H] = fresnel_prop(E0,ps,lam,z)
%Function Input: Initial field in x-y, wavelength lam, pixel size in um,  propagation  distance z
%Function output: Final field in x-y after Fresnel Propagation
%(ref pg 67, J Goodman, Introduction to Fourier Optics)
[M,N] = size(E0);
xsize =  ps*N; ysize = ps*M;

x = linspace(-xsize/2,xsize/2,N);   %Creating the x-y grid
y = linspace(-ysize/2,ysize/2,M);

%Proper way of creating frequency axis
wx =2*pi*(0:(N-1))/N; %Create unshifted default omega axis
fx = 1/ps*(wx-pi*(1-mod(N,2)/N))/2/pi;
%Shift zero to center - for even case, pull back by pi, for odd case by pi(1-1/N)

wy =2*pi*(0:(M-1))/M; %Create unshifted default omega axis
fy = 1/ps*(wy-pi*(1-mod(N,2)/N))/2/pi;
%Shift zero to center - for even case, subtract pi, for odd case subtract pi(1-1/N)

[Fx,Fy] = meshgrid(fx,fy);
```

```matlab
 %Point spread function h=H(kx,ky)  for propagation
H = exp(1i*2*pi/lam*z)*exp(1i*pi*lam*z*(Fx.^2+Fy.^2));
E0fft = fftshift(fft2(E0));
G = H.*E0fft;                    % Convolution in frequency domain
g = ifft2(ifftshift(G));         %Output after deshifting the Fourier transform
Ef=g;                            %Output field ; Intensity is Ef^2
end
```

**Transport of Intensity phase retrieval:**

```matlab
function [Phi_xy,Psi_xy,Grad_Psi_x,Grad_Psi_y,grad2x,grad2y] = tie(IR,IG,IB,ps,lambda,dz,epsilon1,epsilonI)
%Accepts thee intensity matrixes(red, green, blue) :  IR, IG,IB
% z distance between intensity planes :    dz =  del(f)/f *z
%Pixel Size :                              ps (1um)
%Wavelength :                              lambda (0.628)
%epsilon1  :                 Normalize for zero frequency
%epsilonI  :                 Normalize for zero intensity
%Output    :                 Reconstructed Phase

%I0 = I1; %Central plane; can be passed as an argument, here I0~I1
N = size(IR,1);
k = 2*pi/lambda;
xsize =  ps*(N-1); ysize = ps*(N-1); %square grid with N*N points
epsilon = 01e1;
```

```matlab
%Create space
x = linspace(-xsize/2,xsize/2,N);   %N point sampling over xsize
y = linspace(-ysize/2,ysize/2,N);

%Create frequency axis
wx =2*pi*(0:(N-1))/N; %Create unshifted default omega axis
fx = 1/ps*(wx-pi*(1-mod(N,2)/N))/(2*pi); %Shift zero to centre - for even case, pull back by pi, for odd case by pi(1-1/N)
[Fx,Fy] = meshgrid(fx,fx);

%Laplacian
Del2_Psi_xy = (1*k*((IR+IB/2)-IG)/dz);
Psi_xy = poisson_solve_symm(Del2_Psi_xy,ps,N,epsilon1);
[Grad_Psi_x, Grad_Psi_y] = gradient(Psi_xy/ps); %Take the x and y gradients
Grad_Psi_x = Grad_Psi_x./(IG+epsilonI);Grad_Psi_y = Grad_Psi_y./(IG+epsilonI);
%Divide by intensity
[grad2x,dummy1] = gradient(Grad_Psi_x/ps);[dummy2,grad2y]=gradient(Grad_Psi_y/ps);
Del2_Phi_xy = grad2x +grad2y;
Phi_xy = poisson_solve_symm(Del2_Phi_xy,ps,N,epsilon2);
end
```

**Auxilliary function for solving Laplacian  (Poisson Solver)**

```matlab
function Psi_xy = poisson_solve_symm(func,ps,epsilon)
```

```matlab
Del2_Psi_xy = func; %Epsilon added to denominator for divide by zero exception

Del2_Psi_uv = fftshift(fft2(Del2_Psi_xy));

Psi_uv = Del2_Psi_uv./(-4*pi^2*(Fx.^2+Fy.^2+epsilon));

Psi_xy = ifft2(ifftshift(Psi_uv),'symmetric');

end
```

**Data**:

Available upon request from first or last author(s).